\documentclass[]{jfm}

\usepackage{graphicx}
\usepackage[justification=justified]{caption}
\usepackage{newtxtext}
\usepackage{newtxmath}
\usepackage{natbib}
\usepackage{hyperref}
\usepackage{xcolor}
\usepackage{amsmath}
\hypersetup{
    colorlinks = true,
    urlcolor   = blue,
    citecolor  = black,
}

\newcommand{\RomanNumeralCaps}[1]
\linenumbers

\title{Inertial enhancement of the polymer diffusive instability}

\author{Miles M.P. Couchman\aff{1}
  \corresp{\email{mmpc@yorku.ca}},
  Miguel Beneitez\aff{2},
  Jacob Page\aff{3}
 \and Rich R. Kerswell\aff{2}}

\affiliation{\aff{1}Department of Mathematics and Statistics, York University, Toronto, ON M3J 1P3, Canada
\aff{2}Department of Applied Mathematics and Theoretical Physics, University of Cambridge, Cambridge CB3 0WA, UK
\aff{3}School of Mathematics, University of Edinburgh, Edinburgh, EH9 3FD, UK}

\begin{document}
\maketitle

\begin{abstract}

Beneitez et al. ({\em Phys. Rev. Fluids}, {\bf 8}, L101901, 2023) have recently discovered a new linear ``polymer diffusive instability'' (PDI) in inertialess rectilinear viscoelastic shear flow using the FENE-P model when polymer stress diffusion is present. Here, we examine the impact of inertia on the PDI for both plane Couette (PCF) and plane Poiseuille (PPF) flows under varying Weissenberg number $\textit{W}$, polymer stress diffusivity $\varepsilon$, solvent-to-total viscosity ratio $\beta$, and Reynolds number $\textit{Re}$, considering the FENE-P and simpler Oldroyd-B constitutive relations.  Both the prevalence of the instability in parameter space and the associated growth rates are found to significantly increase with $\textit{Re}$. For instance, as $Re$ increases with $\beta$ fixed, the instability emerges at progressively lower values of $W$ and $\varepsilon$ than in the inertialess limit, and the associated growth rates increase linearly with $Re$ when all other parameters are fixed. For finite $Re$, it is also demonstrated that the Schmidt number $Sc=1/(\varepsilon Re)$ collapses curves of neutral stability obtained across various $Re$ and $\varepsilon$. The observed strengthening of PDI with inertia and the fact that stress diffusion is always present in time-stepping algorithms, either implicitly as part of the scheme or explicitly as a stabiliser, implies that the instability is likely operative in computational work using the popular Oldroyd-B and FENE-P constitutive models. The fundamental question now is whether PDI is physical and observable in experiments, or is instead an artifact of the constitutive models that must be suppressed.

\end{abstract}



\section{Introduction}
\label{sec:Intro}

The addition of polymers to a Newtonian solvent can induce dramatically different flow behaviours compared to those observed in the Newtonian fluid alone \citep{datta2022perspectives, sanchez2022understanding}. In industrial processes, for instance, viscous polymer melts are susceptible to instabilities which constrain the maximum extrusion rate \citep{petrie1976instabilities}, while polymer additives are used in oil pipelines to reduce turbulent wall drag \citep{virk1975drag}. Two particularly important viscoelastic phenomena are the existence of `elastic turbulence' (ET), a chaotic flow state sustained in the absence of inertia \citep{groisman2000elastic, steinberg2021elastic}, and  `elasto-inertial turbulence' (EIT), an inherently two-dimensional state arising when both inertia and elasticity are present \citep{samanta2013elasto, sid2018two, choueiri2021experimental}. While the initial pathway to ET in curvilinear geometries is understood \citep{larson1990purely,pakdel1996elastic,shaqfeh1996purely, datta2022perspectives}, relatively little is known about what happens in rectilinear viscoelastic flows.

Initial progress in characterizing ET in rectilinear situations arose through consideration of Kolmogorov flow over a 2-torus, where \cite{boffetta2005viscoelastic} found a linear instability leading to ET \citep{berti2010elastic}. \cite{garg2018viscoelastic} subsequently discovered a centre-mode instability in viscoelastic pipe flow at finite Reynolds number $Re$, which was later also identified in plane Poiseuille flow (PPF, hereafter referred to as channel flow) \citep{khalid2021centre} but notably not in plane Couette flow (PCF). Interestingly, this instability could only be traced down to $Re=0$ in channel flow \citep{khalid2021continuous, buza2022weakly}. The finite-amplitude state resulting from this instability is an `arrowhead' travelling wave \citep{page2020exact,buza2022finite, morozov2022coherent} which has been observed in channel flow EIT \citep{dubief2022first} and, in retrospect, ET in 2D Kolmogorov  flow \citep{berti2010elastic}. In channel flow, efforts have began to establish a dynamical link between the arrowhead solution and both ET \citep{lellep2023linear} and EIT \citep{beneitez2023multistability}. In the latter case in two-dimensions, there does not appear to be a simple dynamical pathway between these arrowhead solutions, where the dynamics is concentrated near the midplane, and EIT \citep{beneitez2023multistability}, which seems more dependent on a near-wall mechanism \citep{shekar2019critical,shekar2021tollmien, dubief2022first}.

The very recent discovery of a new wall-mode ``Polymer Diffusive Instability'' (PDI) in plane Couette flow at $Re=0$ \citep{beneitez2022linear}, however, has added another intriguing possibility for the origin of ET. This instability is dependent on the existence of small but non-vanishing  polymer stress diffusion which is invariably present in any time-stepping algorithm, whether added explicitly to stabilise a numerical scheme like a spectral method (see e.g. \cite{dubief2023elasto}) or arising implicitly such as through a finite difference formulation (see e.g. \cite{zhang2015review,pimenta2017stabilization}). The PDI wall mode is primarily confined to a boundary layer of thickness $\sqrt{\varepsilon}$, where $\varepsilon$ is the (small) diffusion coefficient, traveling at the wall speed with a streamwise wavelength on the order of the boundary layer thickness. The instability is robust to the choice of boundary conditions applied to the polymer conformation equation, and has growth rates which remain $O(1)$ as $\varepsilon \rightarrow 0$. Direct numerical simulations have demonstrated that PDI can lead to a sustained three-dimensional turbulent state, thus providing a potential mechanism for the origin of an ET-like state in the popular finitely-extensible nonlinear elastic constitutive model of Peterlin (FENE-P) \citep{beneitez2022linear}.

While PDI has the potential to be a viscoelastic instability of significant importance, there is an important caveat: the instability emerges at small length scales which can, depending on the parameters, approach the order of the polymer gyration radius \citep{beneitez2022linear}, violating the continuum approximation. There is thus a question of whether the instability is physical or actually an artifact of the Oldroyd-B and FENE-P models. Either possibility has important implications: if the instability is a physical phenomenon then it provides a pathway to ET and EIT, albeit one which will likely be challenging to establish experimentally due to the small length scales involved; {\em or}, it is an artificial feature of the popular FENE-P model, which can compromise its predictions. It thus appears important to now establish the prevalence of PDI across a much wider region of parameter space than was considered in the initial study of \cite{beneitez2022linear}. Therefore, we here map out the regions where PDI is operative at finite $Re$, considering both plane Couette flow (`PCF') and the more experimentally-relevant channel flow scenarios. Both the prevalence of PDI and the associated growth rates are found to significantly increase at finite $Re$ and are relatively insensitive to the bulk flow geometry. PDI is therefore a candidate to trigger both ET {\em and} EIT in simulations using the FENE-P model.

\section {Formulation} \label{sec:Formulation}
We consider the following dimensionless equations governing the flow of an incompressible viscoelastic fluid:
\refstepcounter{equation}
$$
Re\left(\frac{\partial\mathbf{u}}{\partial t}+\left(\mathbf{u}\cdot\nabla\right)\mathbf{u}\right)=-\nabla p+\beta\nabla^{2}\mathbf{u}+\left(1-\beta\right)\nabla\cdot\boldsymbol{\tau},
\quad
\nabla\cdot\mathbf{u}=0,
\eqno{(\theequation{\mathit{a},\mathit{b}})} 
\label{eq:Gov1_NS}
$$
\begin{equation}
\frac{\partial\mathbf{\mathbf{c}}}{\partial t}+\left(\mathbf{u}\cdot\nabla\right)\mathbf{c}+\boldsymbol{\tau}=\mathbf{c}\cdot\nabla\mathbf{u}+\left(\nabla\mathbf{u}\right)^{T}\cdot\mathbf{\mathbf{c}}+\varepsilon\nabla^{2}\mathbf{\mathbf{c}}, \label{eq:Gov2_Conf}
\end{equation}
where $\mathbf{u}=(u,v,w)$ and $p$ denote the velocity and pressure fields, respectively, and $\boldsymbol{\tau}$ denotes the polymeric contribution to the stress tensor. Following \cite{beneitez2022linear}, the equations have been nondimensionalized using the channel half-width $H$ and a characteristic flow speed $U_{0}$, taken to be the wall speed in plane Couette flow or the centerline velocity of the base channel flow. In (\ref{eq:Gov1_NS}a), the Reynolds number $Re:=U_{0} H/\nu_T$ describes the ratio of inertial to viscous forces (with $\nu_T = \nu_S + \nu_P$ denoting the total kinematic viscosity comprised of solvent and polymer components), and $\beta := \nu_S/\nu_T$ denotes the ratio of solvent to total viscosity. The polymeric stress tensor $\boldsymbol{\tau}$ may be described in terms of the polymer orientation through the conformation tensor $\mathbf{c}$ as in (\ref{eq:Gov2_Conf}). We emphasize that the inclusion of a polymer stress diffusion term $\varepsilon\nabla^{2}\mathbf{\mathbf{c}}$ (associated with diffusivity $\varepsilon := \left(ReSc\right)^{-1}$, where $Sc$ denotes the Schmidt number) is the crucial ingredient for the polymer diffusive instability (PDI) identified by \cite{beneitez2022linear}. In the inertialess limit, $\varepsilon=\textit{Sc}^{-1}$ when the governing equations are non-dimensionalized using viscous scales.

To close equations (\ref{eq:Gov1_NS}-\ref{eq:Gov2_Conf}), the FENE-P constitutive model is used to relate $\boldsymbol{\tau}$ and $\mathbf{c}$:
\begin{equation}
\boldsymbol{\tau}:=\frac{f\left(\mathrm{tr}\,\mathbf{c}\right)\mathbf{c}-\mathbf{I}}{W},\ \ f\left(s\right):=\left(1-\frac{s-3}{L^{2}}\right)^{-1}, \label{eq:FENE-P}
\end{equation}
where $\mathbf{I}$ is the identity matrix, $L$ denotes the maximum extensibility of the polymer chains and $W:=U_{0} \lambda /H$, the Weissenberg number, describes the ratio of time-scales for polymer relaxation ($\lambda$) to the flow. In the limit $L\rightarrow \infty$, the simpler Oldroyd-B model is obtained. Inspection of (\ref{eq:Gov1_NS}-\ref{eq:FENE-P}) reveals five parameters of interest governing the flow dynamics: $Re$, $W$, $\beta$, $\varepsilon$, $L$.

We analyze the linear stability of (\ref{eq:Gov1_NS}-\ref{eq:FENE-P}) by perturbing them about their base state: $\mathbf{u}=\mathbf{U}+\mathbf{u}^{*}$, $p=P+p^{*}$, $\boldsymbol{\tau}=\mathbf{T}+\boldsymbol{\tau}^{*}$, $\mathbf{c}=\mathbf{C}+\mathbf{c}^*$. Our coordinate system ($x$, $y$, $z$) is aligned with the streamwise, wall-normal, and spanwise directions of the channel, respectively. The base state  $\left(U(y),C_{xx}(y),C_{yy}(y),C_{zz}(y),C_{xy}(y)\right)$ satisfies
\begin{subeqnarray}   
-\partial_{x}P+\beta U''+\left(1-\beta\right)\partial_{y}T_{xy}=0,\\
f\left(\mathrm{tr}\mathbf{C}\right)\mathbf{C}-\varepsilon W\mathbf{C}''-\begin{pmatrix}2WU'C_{xy}+1 & WU'C_{yy} & 0\\
WU'C_{yy} & 1 & 0\\
0 & 0 & 1
\end{pmatrix}=0,  \label{eq:base}
\end{subeqnarray}
where primes indicates derivatives in the wall-normal ($y$) direction. We use pressure gradients $\partial_{x} P = 0$ and $\partial_{x} P = -2$ for plane Couette and channel flow, respectively. While we here compute the base flow accounting for the presence of a finite diffusivity $\varepsilon$ in (\ref{eq:base}b), it is worth noting that the presence of diffusion in the base flow is not required to induce PDI; the instability still arises if $\varepsilon=0$ in (\ref{eq:base}b) as the inclusion of $\varepsilon \neq 0$ does not significantly change the base state except in the limit of large $\varepsilon = \mathcal{O}(1)$.

\begin{figure}
  \captionsetup{width=1\linewidth}
  \centerline{\includegraphics[width=1\textwidth]{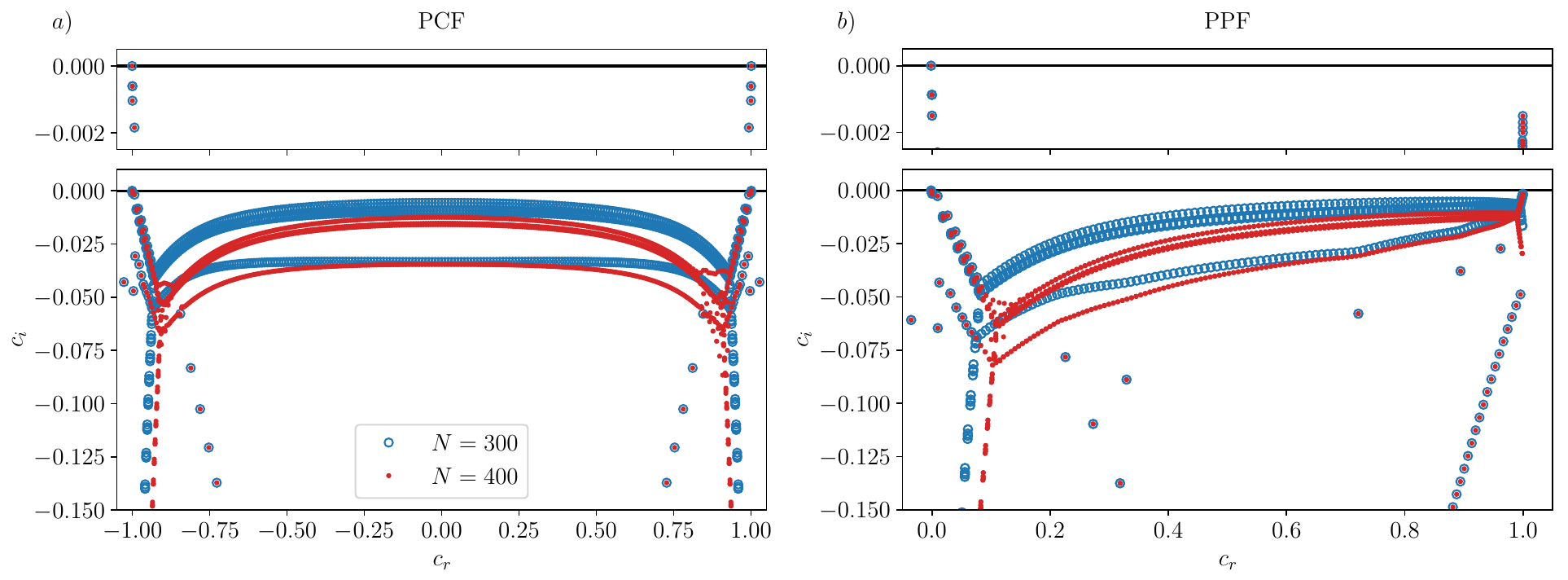}}
  \caption{Eigenvalue spectra, plotted in terms of the real $c_r$ and imaginary $c_i$ components of the complex wavespeed, obtained by solving the system (\ref{eq:PertSys}a-g) using the Oldroyd-B constitutive model at $Re=1000$, $\varepsilon=10^{-5}$ and $\beta=0.9$ for a) plane Couette flow ($W=44.7$, $k=37$) and b) channel flow ($W=22.1$, $k=52$). These parameters are chosen to roughly lie on the corresponding neutral curves plotted in Figure \ref{fig:1_ReVsWi}a. Results using bases of $N=300$ and $400$ Chebyshev polynomials are depicted by blue and red markers, respectively, to demonstrate convergence of the discrete unstable PDI modes highlighted in the zoomed inset above each panel.}
\label{fig:Spectra}
\end{figure} 

Normal mode solutions of the perturbed flow are sought using the ansatz $\phi^{*}\left(x,y,t\right)=\tilde{\phi}\left(y\right)e^{ik\left(x-ct\right)}$, where real-valued $k$ denotes the streamwise wavenumber and $c=c_{r}+ic_{i}$ is a complex wavespeed, with instability arising if $c_i > 0$. The perturbed state is governed by the following system of seven equations for $(\tilde{u},\tilde{v},\tilde{p},\tilde{c}_{xx},\tilde{c}_{yy},\tilde{c}_{zz},\tilde{c}_{xy})$:
\begin{subeqnarray}
    ik\tilde{u}+\tilde{v}' = 0,\ \ \ \ \ \ \\
  Re\left(-ikc\tilde{u}+\tilde{v}U'+ikU\tilde{u}\right)+ik\tilde{p}-\beta\left(-k^{2}\tilde{u}+\tilde{u}''\right)-\left(1-\beta\right)\left(ik\tilde{\tau}_{xx}+\tilde{\tau}'_{xy}\right)=0, \ \ \ \ \ \ \\
  Re\left(-ikc\tilde{v}+ikU\tilde{v}\right)+\ \ \tilde{p}'-\beta\left(-k^{2}\tilde{v}+\tilde{v}''\right)-\left(1-\beta\right)\left(ik\tilde{\tau}_{xy}+\tilde{\tau}'_{yy}\right)=0,\ \ \ \ \ \ \\
    \left[\varepsilon k^{2}+ik\left(U-c\right)\right]\tilde{c}_{xx}+\tilde{v}C'_{xx}+\tilde{\tau}_{xx}-\varepsilon\tilde{c}''_{xx}-2\left(ikC_{xx}\tilde{u}+C_{xy}\tilde{u}'+\tilde{c}_{xy}U'\right)=0,\ \ \ \ \ \ \\
    \left[\varepsilon k^{2}+ik\left(U-c\right)\right]\tilde{c}_{yy}+\tilde{v}C'_{yy}+\tilde{\tau}_{yy}-\varepsilon\tilde{c}''_{yy}\ \ \ \ \ \ \ \ \ \ \ \ \ \ \ -2\left(ikC_{xy}\tilde{v}+C_{yy}\tilde{v}'\right)=0,\ \ \ \ \ \ \\
    \left[\varepsilon k^{2}+ik\left(U-c\right)\right]\tilde{c}_{zz}+\tilde{v}C'_{zz}+\tilde{\tau}_{zz}-\varepsilon\tilde{c}''_{zz}\ \ \ \ \ \ \ \ \ \ \ \ \ \ \ \ \ \ \ \ \ \ \ \ \ \ \ \ \ \ \ \ \ \ \ \ \ \ \ \ \ \ \ \ \ \ \ \ \ \ \ =0,\ \ \ \ \ \ \\
   \left[\varepsilon k^{2}+ik\left(U-c\right)\right]\tilde{c}_{xy}+\tilde{v}C'_{xy}+\tilde{\tau}_{xy}-\varepsilon\tilde{c}''_{xy}\ \ \ \ \ \ -ikC_{xx}\tilde{v}-U'\tilde{c}_{yy}-C_{yy}\tilde{u}'=0.\ \ \ \ \ \ 
    \label{eq:PertSys}
\end{subeqnarray}

By expanding the variables in (\ref{eq:PertSys}) in terms of a basis of Chebyshev polynomials, an eigenvalue problem is obtained which may be solved using standard Python libraries \citep{beneitez2022linear}. A basis of $N=300$ polynomials was used here to target the eigenvalues associated with PDI via an inverse iterations algorithm, which yielded sufficient convergence. Representative eigenvalue spectra are shown in Figure \ref{fig:Spectra} for plane Couette (PCF) and channel (PPF) geometries, illustrating the results using bases of both $N=300$ and $N=400$ Chebyshev polynomials to demonstrate convergence. The unstable PDI mode emerges with characteristic wavespeeds in the vicinity of $\left|c_{r}\right|\approx1$ and $c_r \approx 0$ for PCF and PPF geometries, respectively.

Following \cite{beneitez2022linear}, we choose boundary conditions that set $\varepsilon=0$ at the walls in the governing equations to match the typical configuration used in direct numerical simulations (DNS). While $Sc\sim\mathcal{O}\left(10^{6}\right)$ is characteristic of polymer diffusion in a typical solvent like water (yielding $\varepsilon = 1/(ReSc) \approx 10^{-9}-10^{-6}$ depending on $Re$), in DNS much larger values of $\varepsilon \approx\mathcal{O}(10^{-3})$ are often employed to achieve numerical stability (see e.g. $\S$2.2 of \citet{dubief2023elasto}). Thus, it is commonplace to set $\varepsilon=0$ at the walls in order to recover the idealised limit $\varepsilon\to 0$ \citep{sureshkumar1997direct,samanta2013elasto,dubief2022first}.

\section {Results of linear stability analysis} \label{sec:Results}
The results of our linear stability analysis are presented for both the Oldroyd-B ($\S$\ref{sec:OldroydB}) and FENE-P ($\S$\ref{sec:FENEP}) constitutive relations, delineating the trajectory of neutral curves (characterized by growth rate $\sigma := kc_i=0$) associated with PDI in the four-dimensional parameter space spanned by $(Re, W, \beta,\varepsilon)$. For a given point in parameter space, we perform a sweep over wavenumbers $k$ to ensure that we are tracking the most unstable mode. The results for plane-Couette (PCF) and channel (PPF) geometries are overlaid on the same axes using blue and red colouring, respectively. To promote a collapse of the curves for both geometries, the Weissenberg ($W$) axis is scaled by the respective wall shear rates: $U_{\textrm{wall}}'=\left\{ 1\ (\textrm{PCF}),\ 2\ (\textrm{PPF})\right\}$. As described in $\S$\ref{sec:Intro}, the PDI is a wall mode confined to a boundary layer of thickness $\mathcal{O}(\sqrt{\varepsilon})$ and so its behaviour will primarily be influenced by the wall shear. However, we will demonstrate that this collapse of geometries fails in certain scenarios (such as for large $\varepsilon$), when the boundary layer grows sufficiently large to be influenced by the non-uniform shear profile away from the wall. In the case of finite $Re$, we also demonstrate that curves associated with variable $Re$ and $\varepsilon$ may be collapsed based on $Sc = 1/(\varepsilon Re)$.

\subsection{Oldroyd-B} \label{sec:OldroydB}

%
%
\begin{figure}
  \captionsetup{width=1\linewidth}
  \centerline{\includegraphics[width=1\textwidth]{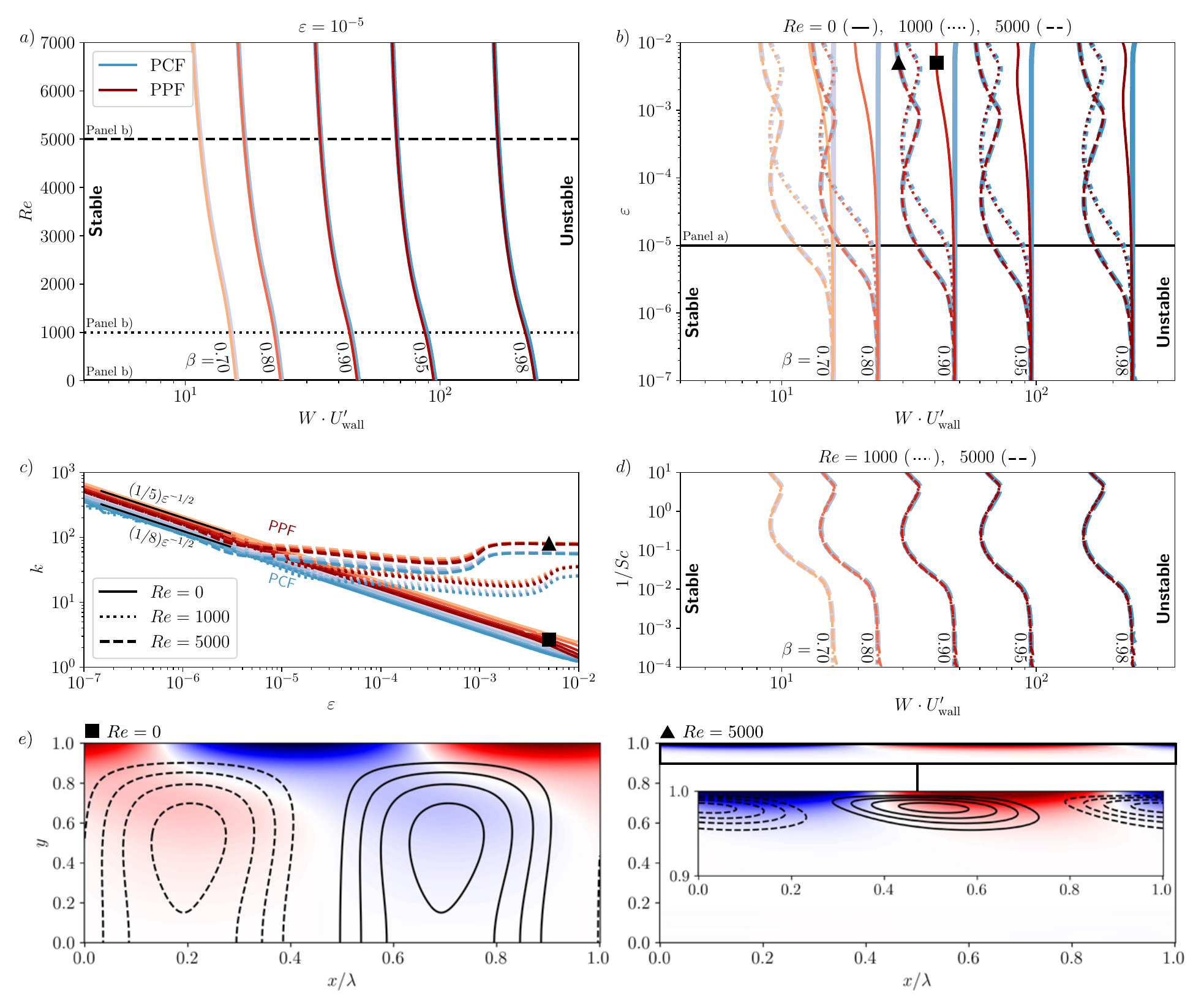}}
  \caption{Curves of neutral stability for plane Couette (`PCF', blue) and channel (`PPF', red) geometries, using the Oldroyd-B constitutive relation for five values of $\beta \in [0.7,0.98]$. a) The $Re$-$W$ plane for fixed $\varepsilon = 10^{-5}$, noting that the PCF and PPF curves are virtually indistinguishable. b) The $\varepsilon$-$W$ plane at three fixed $Re=\left\{ 0,1000,5000\right\}$. In panels a-b, the $W$ axis is scaled by the wall shear rate: $U_{\textrm{wall}}'=\left\{ 1\ (\textrm{PCF}),\ 2\ (\textrm{PPF})\right\} $. c) The streamwise wavenumber $k$ of the PDI, as a function of $\varepsilon$, along each of the neutral curves in panel b. d) A collapse of the $Re=\left\{ 1000,5000\right\}$ curves for both geometries from panel b) based on the inverse Schmidt number $1/Sc = \varepsilon Re$. By plotting $1/Sc$ rather than $Sc$, small $\varepsilon$ occurs at the bottom of panel d) facilitating an easier comparison with panel b). e) Colormap of the trace of the polymer conformation tensor $\mathrm{tr}(\mathbf{c})$ (red and blue denote positive and negative values, respectively), with contours of the streamfunction superimposed, for PPF eigenfunctions in the upper half channel with $\beta=0.9$, at locations indicated by the square and triangular markers in panels b-c. One wavelength $\lambda = 2\pi / k$ of each eigenfunction is shown.}
\label{fig:1_ReVsWi}
\end{figure} 

The trajectory of neutral curves in the $Re$-$W$-$\varepsilon$ volume are presented in Figure \ref{fig:1_ReVsWi} for various $\beta\in\left[0.7,0.98\right]$. Figure \ref{fig:1_ReVsWi}a considers the $Re$-$W$ plane at a fixed $\varepsilon = 10^{-5}$, while Figure \ref{fig:1_ReVsWi}b considers the $\varepsilon$-$W$ plane at three fixed $Re\in\left\{ 0,1000,5000\right\}$. Regions of stability and instability are found to the left and right of each curve, respectively. For a given $\beta$, increasing $Re>0$ is found to promote instability at progressively lower values of both $W$ (Figure \ref{fig:1_ReVsWi}a) and $\varepsilon$ (Figure \ref{fig:1_ReVsWi}b). Notably, in the inertialess limit ($Re=0$), \cite{beneitez2022linear} found that the neutral curves tracked a roughly fixed $W \sim \beta / (1-\beta)$ for $\varepsilon \lesssim 10^{-2}$ (see their Figure 2b) leading to the disappearance of PDI at finite $W$ in the limit $\beta \rightarrow 1$. Here, for $Re>0$, our Figure \ref{fig:1_ReVsWi}b demonstrates that while for small $\varepsilon=\mathcal{O}(10^{-7})$ the neutral curves do indeed still track a constant $W$, they then begin to significantly deviate to lower $W$ beyond $\varepsilon \gtrsim 10^{-6}$, with the $Re=5000$ case deviating at lower $\varepsilon$ than the $Re=1000$ case. Inertial effects thus play a significant role in promoting a greater prevalence of PDI in the parameter space. In Figure \ref{fig:1_ReVsWi}d, we demonstrate that the $Re=\left\{ 1000,5000\right\}$ curves plotted in Figure \ref{fig:1_ReVsWi}b may be collapsed for both geometries based on the Schmidt number $Sc = 1/(\varepsilon Re)$.

The streamwise wavenumbers $k$ associated with the neutral curves in Figure \ref{fig:1_ReVsWi}b are plotted in Figure \ref{fig:1_ReVsWi}c as a function of $\varepsilon$. At $Re=0$, $k$ follows the $1/\sqrt{\varepsilon}$ scaling reported by \cite{beneitez2022linear} for all $\varepsilon$, whereas for $Re>0$, $k$ deviates significantly from this scaling to plateau to a roughly constant value for $\varepsilon \gtrsim 10^{-5}$, with the unstable modes at higher $Re$ being more tightly confined to the wall (as indicated by a larger $k$). This deviation from the $1/\sqrt{\varepsilon}$ scaling for $Re>0$ corresponds to the previously noted deviation of the neutral curves away from a constant $W$ in Figure \ref{fig:1_ReVsWi}b, where the curves begin turning to lower $W$ in the vicinity of $\varepsilon \approx 10^{-6}-10^{-5}$. The behaviour of $k$ in Figure \ref{fig:1_ReVsWi}c also explains the mismatch in the collapse of the PCF and PPF curves in Figure \ref{fig:1_ReVsWi}b at higher $\varepsilon$ for $Re=0$ (see solid lines). Specifically, at higher $\varepsilon$, $k$ becomes sufficiently small at $Re=0$ such that the instability is no longer strongly confined to the wall (see square eigenfunction, Figure \ref{fig:1_ReVsWi}e) and so the local wall shear, $U'_\textrm{wall}$, does not accurately describe the non-uniform shear profile influencing the instability. Conversely, the curves for $Re=\left\{ 1000,5000\right\} $ do remain collapsed at high $\varepsilon$, as the wavenumbers $k$ in Figure \ref{fig:1_ReVsWi}c plateau to sufficiently large values such that the instability remains confined to the wall (see triangle eigenfunction, Figure \ref{fig:1_ReVsWi}e). In Figure \ref{fig:1_ReVsWi}c, it is also worth noting that the prefactor of the PPF scaling (1/5) is roughly twice that of the PCF scaling (1/8), thus indicating that PDI is more tightly confined to the wall in the channel geometry. 

\subsection{FENE-P} \label{sec:FENEP}

\begin{figure}
  \captionsetup{width=1\linewidth}
  \centerline{\includegraphics[width=1\textwidth]{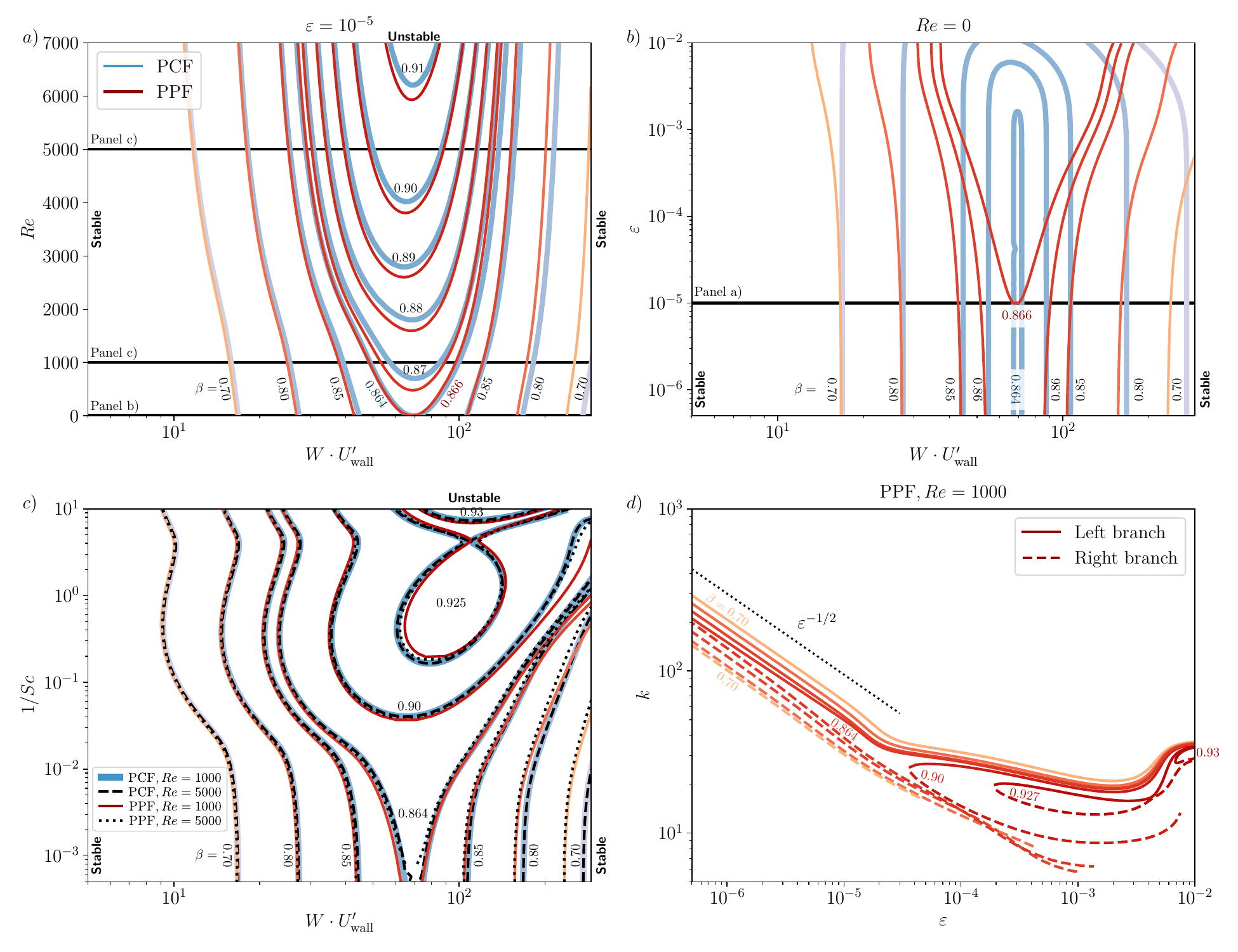}}
  \caption{Curves of neutral stability using the FENE-P constitutive relation with a fixed extensibility $L=200$, presented analogously to the Oldroyd-B curves in Figure \ref{fig:1_ReVsWi} for various $\beta$. Curves are shown in a) the $Re$-$W$ plane with a fixed $\varepsilon=10^{-5}$, b) the $\varepsilon$-$W$ plane for fixed $Re=0$, and c) the $1/Sc = \varepsilon Re$ versus $W$ plane, which collapses curves obtained at $Re=1000$ and $5000$ for both geometries. By plotting $1/Sc$ rather than $Sc$, small $\varepsilon$ occurs at the bottom of panel c) facilitating an easier comparison with panel b). Panel d) illustrates the dependence of the optimal streamwise wavenumber $k$ on $\varepsilon$ for the PPF curves in panel c) at $Re=1000$. Left and right branches of the curves in panel c) are distinguished here using solid and dashed lines, respectively. Comparison with the Oldroyd-B curves in Figure \ref{fig:1_ReVsWi}c reveals that the left branches behave similarly to the single Oldroyd-B branch, deviating from the $1/\sqrt{\varepsilon}$ scaling at large $\varepsilon$, while the right branches retain this scaling to the highest $\varepsilon$ considered.}
\label{fig:3_FENEP}
\end{figure}

We now consider how a finite polymer extensibility $L$ modifies the behaviour of the instability as compared to the Oldroyd-B case ($L\rightarrow \infty$) presented in $\S$\ref{sec:OldroydB}. Focusing first on $L=200$, Figure \ref{fig:3_FENEP} illustrates the much richer behaviour of the neutral curves for the FENE-P case, presented in an analogous manner to Figures \ref{fig:1_ReVsWi}a-d. In the $Re$-$W$ plane (Figure \ref{fig:3_FENEP}a), a finite $L$ introduces two notable differences compared to Oldroyd-B. First, the neutral curves for a given $\beta$ now have a left and right branch and so the range of instability is bounded by an upper value of $W$. Second, there is now a critical value of $\beta$ ($\approx 0.865$ for both geometries) above which the neutral curves no longer intersect the $Re=0$ axis, at fixed $\varepsilon=10^{-5}$. Therefore, inertial effects are once again demonstrated to promote PDI, generating instability at finite $Re$ for ultra-dilute polymer solutions ($\beta \rightarrow 1$) that would otherwise remain stable at $Re=0$.

Neutral curves in the $\varepsilon$-$W$ plane for the inertialess case $Re=0$ are shown in Figure \ref{fig:3_FENEP}b, exhibiting notable differences to the Oldroyd-B $Re=0$ curves presented in Figure \ref{fig:1_ReVsWi}b. While for Oldroyd-B the plane Couette and channel curves adopt roughly the same shape, the FENE-P curves display dramatically different behaviours at large $\varepsilon$. Specifically, the plane Couette curves have an inverted `U'-shape, highlighting that the instability ceases to exist at large $\varepsilon$ for certain $\beta$ (e.g. see the $\beta=0.86$ curve which reaches a maximum value at $\varepsilon \approx 6\times 10^{-3}$). Conversely, the channel curves are roughly `U'-shaped, and the range of instability only increases with increasing $\varepsilon$. 

%
%

At finite $Re$, as for the Oldroyd-B curves in Figure \ref{fig:1_ReVsWi}d, the FENE-P curves shown in Figure \ref{fig:3_FENEP}c at $Re=\left\{ 1000,5000\right\}$ are again found to collapse for both geometries based on the Schmidt number $Sc=1/(\varepsilon Re)$. Comparing Figures \ref{fig:1_ReVsWi}d and \ref{fig:3_FENEP}c reveals two main differences in the structure of the FENE-P and Oldroyd-B curves. First, for $\beta \approx 0.925$, in both geometries a pinch-off phenomenon occurs for the FENE-P case where the neutral curves form a ``bubble'' of instability within the $\varepsilon-W$ plane in an otherwise stable region. Second, in the limit of $\varepsilon \rightarrow 0$ ($Sc \rightarrow \infty$), the neutral curves behave differently as compared to the $Re=0$ case reported by \cite{beneitez2022linear}. In particular, there now appears to be a critical value of $\beta \approx 0.865$ (note the similarity to the critical $\beta$ in Figure \ref{fig:3_FENEP}a corresponding to lift-off from the $Re=0$ axis) at which the two branches of the neutral curve asymptotically approach each other in the limit of small $\varepsilon$. Curves associated with $\beta$ greater than this critical value are thus observed to turn back at higher $\varepsilon$ (see e.g. the $\beta=0.9$ curves, Figures \ref{fig:3_FENEP}c), suggesting that, at finite $Re$, PDI will not exist in the limit $\varepsilon \rightarrow 0$ for all $\beta$. We note another possibility, however, which is that an `hourglass'-like pinch-off behaviour occurs, in which the critical curves touch at some finite $\varepsilon$ (appearing here to be around $Sc \approx 10^{4}$), but then separate again for lower $\varepsilon$. The concave-up neutral curves (e.g. $\beta=0.9$) seen here, might then be reflected as concave-down branches at much lower $\varepsilon$ and instabilities for such $\beta$ could then still exist in the $\varepsilon \rightarrow 0$ limit. As $\varepsilon \rightarrow 0$ isolates the instability to an increasingly thin boundary layer at the wall, significantly increased computational power is required to resolve the neutral curves for $Sc > 10^{4}$ and so we do not further consider this behaviour here. It is also worth noting in Figures \ref{fig:3_FENEP}c that while increasing $Re$ does not significantly influence the horizontal position of the neutral curves along the $W$ axis, the increase in inertial effects does induce a significant downward shift of the curves in $\varepsilon$ (by a factor proportional to $Re$), thus increasing the prevalence of PDI in parameter space. 

%
%
\begin{figure}
  \captionsetup{width=1\linewidth}
  \centerline{\includegraphics[width=1\textwidth]{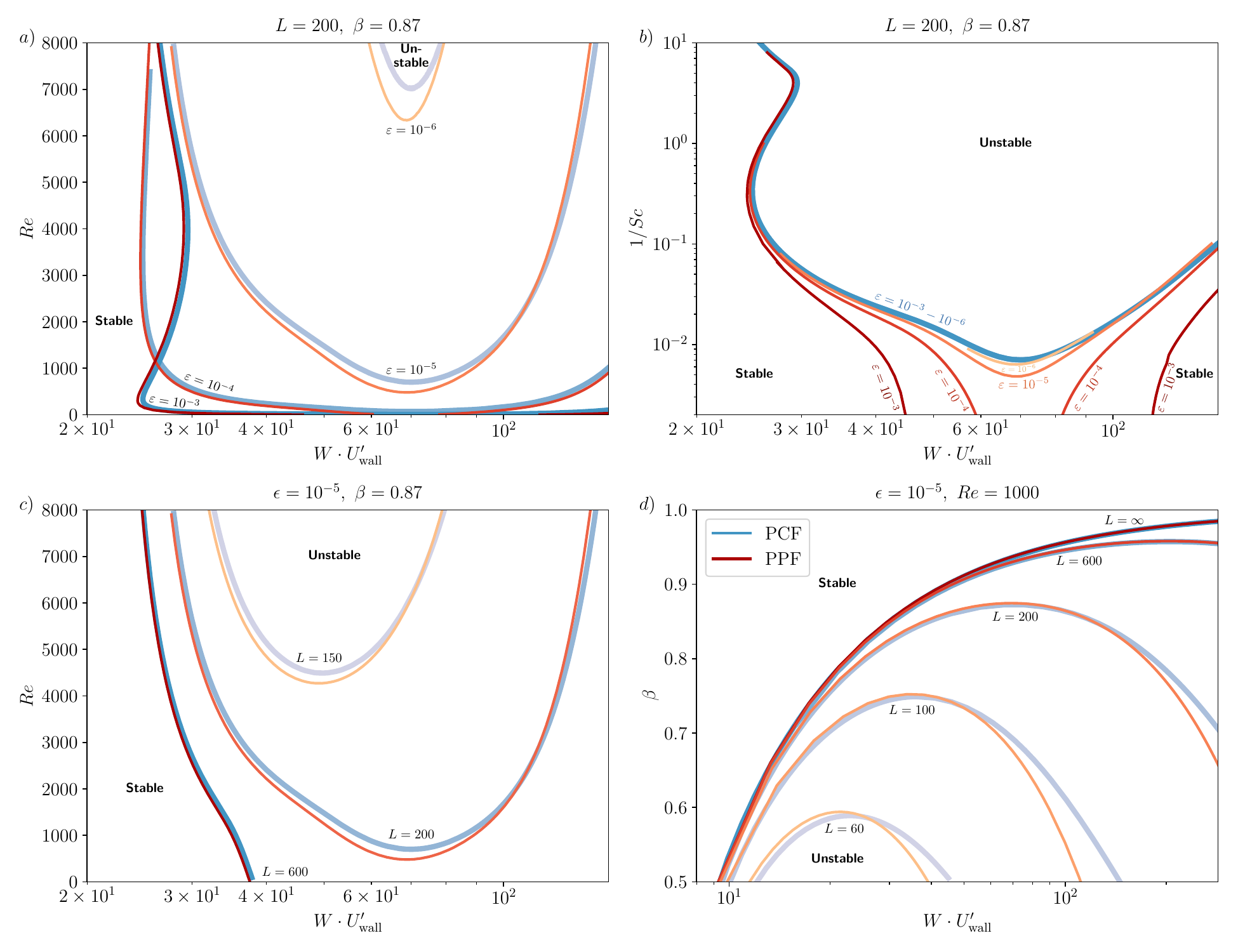}}
  \caption{a) Dependence of the $\beta=0.87$ neutral curve presented in Figure \ref{fig:3_FENEP}a on variable $\varepsilon$, with fixed $L=200$. b) A collapse of the curves in panel a) based on the inverse Schmidt number $1/Sc = \varepsilon Re$, plotted analogously to Figures \ref{fig:1_ReVsWi}d and \ref{fig:3_FENEP}c such that small $\varepsilon$ appears at the bottom of the plot. The plane Couette curves (PCF) exhibit a perfect collapse and are virtually indistinguishable, whereas the $\varepsilon = 10^{-3}$ and $10^{-4}$ channel curves intersect the $Re=0$ axis in panel a) and thus diverge to infinite $Sc$ in panel b). c) Dependence of the $\beta=0.87$ neutral curve presented in Figure \ref{fig:3_FENEP}a on variable $L$, with fixed $\varepsilon = 10^{-5}$. d) Neutral curves in the $\beta$-$W$ plane for fixed $\varepsilon=10^{-5}$ and $Re=1000$.}
\label{fig:9_VariableLNew}
\end{figure} 

The streamwise wavenumbers $k$ associated with the FENE-P channel (PPF) neutral curves at $Re=1000$ in Figure \ref{fig:3_FENEP}c are presented in Figure \ref{fig:3_FENEP}d as a function of $\varepsilon$, to compare with the scaling of the Oldroyd-B curves in Figure \ref{fig:1_ReVsWi}c. As for Oldroyd-B, the left branches follow the $1/\sqrt{\varepsilon}$ scaling reported by \cite{beneitez2022linear} for $\varepsilon \lesssim 10^{-5}$, when the neutral curves are roughly independent of $W$ in Figure \ref{fig:3_FENEP}c. In contrast, the right branches follow this $k$ scaling for the entire range of $\varepsilon$ considered here, thus explaining the slight mismatch in the collapse of the right branches of the two geometries at low $\varepsilon$ in Figure \ref{fig:3_FENEP}c (see e.g. the right branches of the $\beta=0.7$ and 0.8 curves); in this regime, $k$ has become sufficiently small such that the instability is no longer confined to the wall and thus the wall shear $U_{\textrm{wall}}'$ is not entirely suitable for scaling the $W$ axis. In Figure \ref{fig:3_FENEP}d, we also note that for $\beta \gtrsim 0.864$, the pinchoff value seen in Figure \ref{fig:3_FENEP}c at $Sc = \mathcal{O}(10^{4})$, the neutral curves turn back to higher $\varepsilon$ before the $1/\sqrt{\varepsilon}$ scaling regime is reached.

In Figure \ref{fig:9_VariableLNew}, we further consider the behaviour of the neutral curves presented in Figure \ref{fig:3_FENEP} by varying parameters that were previously held fixed. While the $Re$-$W$ plane was considered with a fixed $\varepsilon = 10^{-5}$ in Figure \ref{fig:3_FENEP}a, in Figure \ref{fig:9_VariableLNew}a we focus on the $\beta=0.87$ neutral curve and demonstrate its dependence on variable $\varepsilon$, where it is found that decreasing $\varepsilon$ pushes the region of PDI to progressively higher $Re$. In Figure \ref{fig:9_VariableLNew}b, we collapse the curves from Figure \ref{fig:9_VariableLNew}a based on the Schmidt number $Sc = 1/(\varepsilon Re)$. A near perfect collapse is observed for the plane Couette (PCF) curves, while some channel (PPF) curves diverge to $Sc \rightarrow \infty$ due to their intersection with the $Re=0$ axis in Figure \ref{fig:9_VariableLNew}a. Additionally, having only considered a fixed polymer extensibility $L=200$ in Figure \ref{fig:3_FENEP}, in Figures \ref{fig:9_VariableLNew}c,d we consider the influence of variable $L$ on the neutral curves in the $Re$-$W$ and $\beta$-$W$ planes, respectively. In Figure \ref{fig:9_VariableLNew}c, decreasing $L$ induces a turn-back of the neutral curves at progressively higher values of $Re$, decreasing the extent of PDI in the $Re$-$W$ plane. Figure \ref{fig:9_VariableLNew}d, at fixed $Re=1000$, displays the same qualitative features as those reported in the inertialess limit by \cite{beneitez2022linear} (see their Figure 2c), again demonstrating that decreasing $L$ decreases the prevalence of PDI in the $\beta$-$W$ plane.\\

\subsection{Growth Rates} \label{sec:growthRates}

\begin{figure}
  \captionsetup{width=1\linewidth}
  \centerline{\includegraphics[width=1\textwidth]{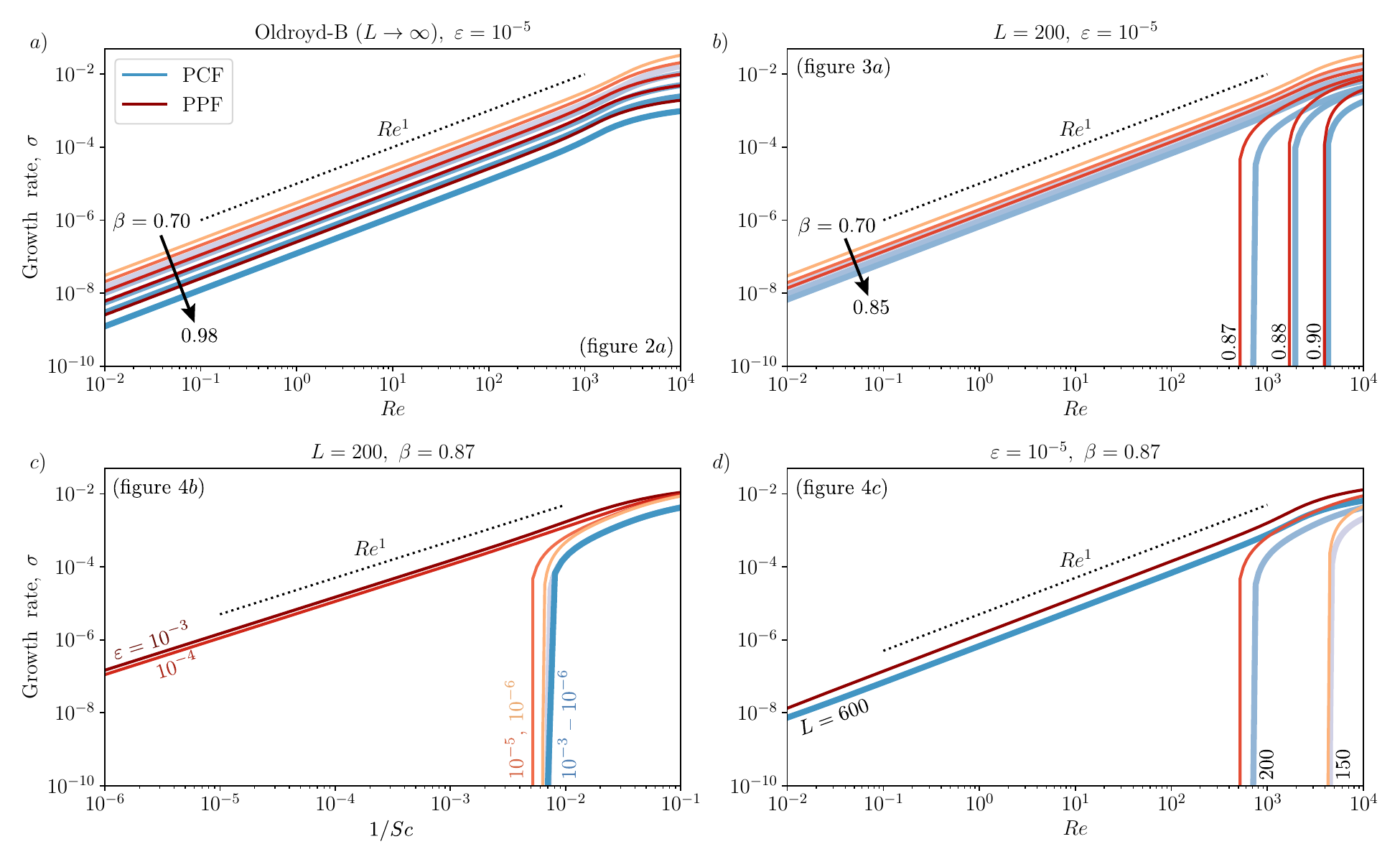}}
  \caption{Growth rates $\sigma := kc_{i}$ of the most unstable mode as a function of Reynolds number $Re$ obtained at a fixed $W$ corresponding to either the intersection of the neutral curve with the $Re=0$ axis or the minimum of the neutral curve in $Re$ for cases in which there is no intersection with $Re=0$. Growth rates are shown for neutral curves spanning various $\beta$ in a) figure \ref{fig:1_ReVsWi}a (Oldroyd-B, fixed $\varepsilon=10^{-5}$) and b) figure \ref{fig:3_FENEP}a (fixed $L=200$, $\varepsilon=10^{-5}$), and neutral curves at fixed $\beta=0.87$ spanning c) various $\varepsilon$ in figure \ref{fig:9_VariableLNew}b (fixed $L=200$) and d) various $L$ in figure \ref{fig:9_VariableLNew}c (fixed $\varepsilon=10^{-5}$).}
\label{fig:2_Growth}
\end{figure} 

While we have thus far considered the behaviour of the neutral curves associated with PDI, it is also informative to consider how the growth rate of PDI evolves with $Re$ in regions of instability. Starting with each $\beta$ curve in Figure \ref{fig:1_ReVsWi}a (Oldroyd-B), we first fix the value of $W$ at which the neutral curve intersects the $Re=0$ axis, and where the growth rate is thus zero by definition. At this fixed $W$, we then increase $Re$ incrementally and track the growth rate of the most unstable PDI mode, as shown in Figure \ref{fig:2_Growth}a. In Figure \ref{fig:2_Growth}b, we repeat this process for the FENE-P curves presented in Figure \ref{fig:3_FENEP}a. As some FENE-P curves do not intersect the $Re=0$ axis, in these cases we fix $W$ to correspond to the minimum $Re$ of the neutral curve, and then increase $Re$ from there. In Figures \ref{fig:2_Growth}c-d, we again repeat this process for the neutral curves plotted in Figures \ref{fig:9_VariableLNew}c-d, respectively.

In all cases, for neutral curves that intersect the $Re=0$ axis, the growth rate is observed to grow linearly with $Re$ as one moves away from the neutral curve, emphasizing the intensification of PDI due to the presence of inertia. Notably, the streamwise wavenumber $k$ remains virtually constant during this scaling, until $Re \gtrsim 10^3$ at which point the most unstable $k$ beings to vary and the linear scaling breaks down. For the subset of FENE-P curves that do not intersect the $Re=0$ axis (e.g. $\beta=0.87$, 0.88, 0.90 in Figure \ref{fig:9_VariableLNew}b), it is also intriguing that the growth rates display a dramatic increase with $Re$ to quickly join the linear $Re$ scaling of the curves that do intersect the $Re=0$ axis. We note that the relative vertical translation of the various curves in Figure \ref{fig:2_Growth} is due to differences in slope between the neutral curves at their intersection with the $Re=0$ axis; at a fixed $W$, these differences in slope will govern the rate at which one moves away from the neutral curve as $Re$ is increased, and hence the observed difference in the growth rate magnitudes.

\vspace{0.5cm}

\section {Conclusions} \label{sec:Discussion}
In this study, we have demonstrated that the polymer diffusive instability (PDI) is active in both plane Couette and channel flows with or without inertial effects, and that the instability intensifies with increasing Reynolds number $Re$. Through exploration of a variety of dimensionless parameters, we have found that PDI is operational across large regions of the parameter space including those relevant to many prior experiments \citep{choueiri2018exceeding,qin2019flow,choueiri2021experimental,jha2021elastically}. In particular, increasing $Re$ enhances the prevalence of the instability, promoting instability at progressively smaller values of both $W$ and $\varepsilon$ than in the inertialess limit. Our results therefore significantly extend the conclusion of \citet{beneitez2022linear} that PDI could also present a possible transition mechanism to EIT as well as ET in FENE-P fluids.

The eigenfunction for PDI is a wall mode, confined to a boundary layer of thickness $\sqrt{\varepsilon}$. As a result, the neutral curves for Couette and channel flow are found to nearly overlap in most regions of parameter space when $W$ is scaled by the wall-shear rate. This collapse breaks down when the streamwise wavenumber $k$ approaches $\mathcal{O}(1)$, as the instability is no longer confined to the wall and thus feels a non-monotonic shear profile in the channel's interior, as occurs for large $\varepsilon$ at $Re=0$ (but notably not at higher $Re$, see Figure \ref{fig:1_ReVsWi}c), and small $\varepsilon$ at high $W$ in FENE-P fluids. The finite extensibility of the polymer chains ($L$) is also found to have a significant impact on the prevalence of PDI, as compared to that predicted by the Oldroyd-B model. Considering $L=200$, we found that for sufficiently high $\beta$, the instability is suppressed at $Re=0$ and only appears at progressively larger $Re$ (Figure \ref{fig:3_FENEP}a). Similarly, beyond a critical value of $\beta$, the instability may also be suppressed at small values of $\varepsilon$ (Figure \ref{fig:3_FENEP}c). Given that PDI emerges as a wall mode, we have also confirmed its presence in cylindrical pipe flow as well as Taylor-Couette flow.

The lengthscale associated with PDI raises some intriguing questions. As indicated by \citet{beneitez2022linear}, PDI can emerge at a lengthscale roughly on the order of the polymer gyration radius, where the continuum assumptions of the FENE-P model may not hold. It is thus possible that PDI is an unphysical feature of the widely-used FENE-P model, which would have significant implications for computational studies of ET, EIT and polymer drag reduction using this model. Until now, such studies may have been unknowingly influenced by PDI, due to the ubiquity of stress diffusion in numerical schemes, either introduced explicitly as a regularisation term or arising implicitly through the discretization scheme. Assessing the relevance of PDI to real viscoelastic fluids is now a key challenge to be confronted.

\backsection[Funding]{We are grateful for the support of EPSRC under grant EP/V027247/1.}

\backsection[Declaration of interests]{The authors report no conflict of interest.}

\bibliographystyle{jfm}
\bibliography{jfm}

\end{document}